\documentclass[a4paper,alpha-refs]{eSpectra}

\journal{aej}

\usepackage{graphicx}
\usepackage{siunitx}
\usepackage[spanish, english]{babel} 
\usepackage{amssymb}
\usepackage{comment}
\usepackage{widetext}
\usepackage{listings}
\usepackage{float}

\usepackage{academicons}
\usepackage{aas-macros}
\usepackage{doi}
\usepackage{fontawesome5}
\usepackage[left]{lineno}

\definecolor{orcidlogocol}{rgb}{0.65, 0.807, 0.223}
\newcommand{\orcid}[1]{$\,$\href{https://orcid.org/#1}{\textcolor{orcidlogocol}{\faOrcid}}}

\usepackage{tikz}
\usetikzlibrary{arrows.meta}
\definecolor{detfill}{HTML}{E8F5EE}\definecolor{detline}{HTML}{2E7D5E}
\definecolor{bhfill}{HTML}{FAE8E4}\definecolor{bhline}{HTML}{A03020}
\definecolor{accfill}{HTML}{FDF0DA}\definecolor{accline}{HTML}{8A5200}
\definecolor{intfill}{HTML}{EDEAFC}\definecolor{intline}{HTML}{4A40A0}
\definecolor{evfill}{HTML}{F2F2EE}\definecolor{evline}{HTML}{666666}
\definecolor{outfill}{HTML}{E8F5E2}\definecolor{outline}{HTML}{3A6E20}
\definecolor{notefill}{HTML}{FAFAFA}\definecolor{noteline}{HTML}{AAAAAA}
\definecolor{seccol}{HTML}{888888}
\newcommand{\arctitle}[1]{{\large\bfseries #1}}
\newcommand{\arcsub}[1]{{\normalsize\itshape #1}}
\newcommand{\arcmath}[1]{{\normalsize #1}}
 
\title{TARTARUS: A High-Performance Python Code for Ray Tracing in Curved Spacetimes}

\author[1,\authfn{1}]{David F. Bambague}
\author[1,\authfn{2}]{Alexis Larrañaga}

\affil[1]{Observatorio Astronómico Nacional, Universidad Nacional de Colombia, Bogotá D.C., Colombia}

\authnote{\authfn{1} Master of Science Student - Astronomy (dbambague@unal.edu.co)}
\authnote{\authfn{2}Associate professor  (ealarranaga@unal.edu.co )}

\papercat{Scientific Article}

\runningauthor{D. Bambague-E. Larrañaga}

\jvolume{2}
\jnumber{1}
\jyear{2024}

\begin{document}

\begin{frontmatter}
\maketitle

\selectlanguage{english}
\begin{abstract}
\justifying
The recent observations of the supermassive black holes M87* and Sagittarius A* by the Event Horizon Telescope collaboration (EHT) have shown the need for robust, efficient, and accessible numerical tools to model light propagation in strongly curved spacetimes. Comparing theoretical accretion models with observational data relies heavily on ray-tracing algorithms that calculate the trajectories of photons traveling from a source plasma to a distant observer. In this paper, we present TARTARUS (Tracer for Astrophysical Ray Trajectories Around Relativistic Ultra-compact Sources), a new, highly modular, and extensible Python-based framework for computing null geodesics around compact objects. By leveraging Just-In-Time (JIT) compilation via Numba, TARTARUS bridges the gap between Python's high-level accessibility and the execution speed of compiled languages like C++ or FORTRAN. The code supports multiple native solvers (including adaptive embedded Runge--Kutta pairs, a Bulirsch--Stoer extrapolator, and a Verlet scheme),  event-detection mechanics for handling structural intersections (e.g. event horizons and accretion disks), and support for both analytical and numerically generated background metrics. We demonstrate the code's physical accuracy and computational efficiency through some tests, including the evaluation of the Hamiltonian constraint conservation and ray-tracing of the shadow and the Novikov-Thorne thin accretion disks around a Kerr black hole.
\end{abstract}

\qquad\quad\textbf{Keywords:} Black Holes - Geodesics - Ray Tracing - Numerical methods.

\selectlanguage{spanish} 
\begin{abstract}
\justifying
Las recientes observaciones de los agujeros negros supermasivos M87* y Sagitario A* realizadas por la colaboración Event Horizon Telescope (EHT) han puesto de manifiesto la necesidad de contar con herramientas numéricas robustas, eficientes y accesibles para modelar la propagación de la luz en espacios fuertemente curvados. La comparación de modelos teóricos de acreción con los datos observacionales depende en gran medida de algoritmos de trazado de rayos que calculan las trayectorias de los fotones que viajan desde un plasma emisor hasta un observador distante. En este artículo presentamos TARTARUS (Tracer for Astrophysical Ray Trajectories Around Relativistic Ultra-compact Sources), un nuevo código en Python, altamente modular y extensible, para el cálculo de geodésicas nulas alrededor de objetos compactos. Al aprovechar la compilación Just-In-Time (JIT) mediante Numba, TARTARUS llena la brecha entre la accesibilidad de alto nivel de Python y la velocidad de ejecución de lenguajes compilados como C++ o FORTRAN. El código es compatible con múltiples métodos de integración propios (incluidos Runge--Kutta de paso adaptativo, un extrapolador de Bulirsch--Stoer y un esquema de Verlet), mecánicas de detección de eventos para manejar intersecciones estructurales (por ejemplo, horizontes de eventos y discos de acreción) y admite tanto métricas analíticas como métricas numéricas. Demostramos la precisión de los parámetros físicos y la eficiencia computacional del código mediante diversas pruebas, incluyendo la evaluación de la conservación del hamiltoniano y el trazado de rayos aplicado a la obtención de la sombra y de la imagen de discos  de acreción delgados de Novikov-Thorne alrededor de un agujero negro de Kerr.
\end{abstract}

\begin{skeywords}
Agujeros Negros - Geodésicas - Trazado de Rayos - Métodos Numéricos
\end{skeywords}
\end{frontmatter}

\selectlanguage{english}

\section{Introduction}

The era of precision strong-field gravity observation was inaugurated by the direct imaging of the black hole shadows of M87* and Sgr A* [\cite{eht2019m87, eht2022sgra}]. The observed structures in the photographies show the macroscopic signatures of gravitational lensing and extreme redshift acting on the radiation emitted by the surrounding plasma in an accretion disk. To interpret these observations, astrophysicists rely on General Relativistic Magnetohydrodynamic (GRMHD) simulations coupled with General Relativistic Radiative Transfer (GRRT) codes.

At the core of these GRRT tools is the integration of null geodesics, which correspond to the paths that light follows through the curved spacetime. While legacy ray-tracing codes have traditionally been developed in C, C++, or FORTRAN (e.g. OSIRIS [\cite{velasquez2022osiris}] or GYOTO [\cite{vincent2011gyoto}]) to maximize computational efficiency, the scientific community has increasingly adopted Python for its vast ecosystem, readability, and ease of modification, although it is clear that the interpreted nature of Python make it too slow for the heavy iterative calculations required in massive ray-tracing grids.

In this paper, we introduce TARTARUS (Tracer for Astrophysical Ray Trajectories Around Relativistic Ultra-compact Sources), a code developed by the research group on Computational Astrophysics at the Observatorio Astronómico Nacional of the Universidad Nacional de Colombia. TARTARUS implements a highly parallelizable and JIT-compiled architecture using the Numba library [\cite{lam2015numba}], allowing a C-like performance while maintaining a pure Python, object-oriented user interface. The code is structured to easily integrate alternative theories of gravity (e.g. hairy black holes or modified gravity models) alongside standard Schwarzschild and Kerr metrics.

We outline the mathematical formalism employed by TARTARUS in Section 2, the detailed computational architecture and integration schemes is shown in Section 3 and the physical validations and performance benchmarks are presented in Section 4. We close with the concluding remarks and some future projects in Section 5.

\section{Mathematical Formalism}
\subsection{The Geodesic Equations}

The propagation of light in a curved background geometry, described by a metric tensor $g_{\mu\nu}$, is governed by the geodesic equation. For numerical integration, it is computationally advantageous to express this system using the Hamiltonian formulation. Hence, we introduce the Hamiltonian function for a test particle with a four-momentum $p_\mu$ as
\begin{equation}
    \mathcal{H} = \frac{1}{2} g^{\mu\nu} p_\mu p_\nu. \label{eq:Hamiltonian}
\end{equation}

The rest mass of photons is zero, which implies the constraint $\mathcal{H} = 0$. The equations of motion, parameterized by an affine parameter $\lambda$, are
\begin{align}
    \frac{dx^\mu}{d\lambda} = &\frac{\partial \mathcal{H}}{\partial p_\mu} = g^{\mu\nu} p_\nu\\
    \frac{dp_\mu}{d\lambda} = &- \frac{\partial \mathcal{H}}{\partial x^\mu} = -\frac{1}{2} (\partial_\mu g^{\alpha\beta}) p_\alpha p_\beta.
\end{align}

TARTARUS uses these equations as an 8-dimensional phase-space array, $q = [t, r, \theta, \phi, p_t, p_r, p_\theta, p_\phi]$, and computes their derivatives with respect to $\lambda$ using highly optimized Numba kernels.

\subsection{The Observer and the Image Plane}

To create an image of the black hole, TARTARUS employs the backward ray-tracing method in which photons are initialized at the detector (the observer's image plane) and traced backward in time toward the strong-field region. The observer is located at a large distance $D$ from the black hole and at an inclination angle $\iota$ with respect to the $z$-axis (rotation axis of the black hole). The celestial coordinates $(\alpha, \beta)$ are introduced in the image plane of the observer to map the pixels of the detector and the initial 4-momentum of a photon at the pixel $(\alpha, \beta)$ is determined by projecting the local basis of the zero-angular-momentum observer (ZAMO) into the global coordinate system of the background metric, following the formalism established by Cunningham [\cite{cunningham1975effects}].

Using the orthonormal tetrad $e_{(a)}^{\mu}$ associated with the observer, the components of the photon's four-momentum in the local frame $p^{(a)}$ are parameterized by the celestial coordinates as
\begin{equation}
    \begin{cases}
    p^{(t)} = &1\\
    p^{(r)} = &\left( 1 - \frac{\alpha^2 + \beta^2}{D^2} \right)^{1/2}\\
    p^{(\theta)} = &\frac{\beta}{D}\\
    p^{(\phi)} = &-\frac{\alpha}{D}.
    \end{cases}
\end{equation}

These local components are then transformed into the global coordinate basis $p^{\mu} = e_{(a)}^{\mu} p^{(a)}$ using the ZAMO tetrad evaluated at the position of the observer, $(t=0, r=D, \theta=\iota, \phi=0)$,
\begin{equation}
\begin{cases}
    p^t = &e_{(t)}^t p^{(t)}\\
    p^r = &e_{(r)}^r p^{(r)}\\
    p^\theta = &e_{(\theta)}^\theta p^{(\theta)}\\
    p^\phi = &e_{(t)}^\phi p^{(t)} + e_{(\phi)}^\phi p^{(\phi)}.
\end{cases}
\end{equation}

For a general metric of the form 
\begin{equation}
    ds^2 = g_{tt}dt^2 +2g_{t\phi}dtd\phi+ g_{rr}dr^2 + g_{\theta \theta}d\theta^2 + g_{\phi \phi}d\phi^2,
\end{equation}
the non-zero tetrad components are 
\begin{align}
    e_{(t)}^t = &1/\sqrt{-g_{tt} + \omega^2 g_{\phi\phi}}\\
    e_{(r)}^r = &1/\sqrt{g_{rr}}\\
    e_{(\theta)}^\theta = &1/\sqrt{g_{\theta\theta}}\\
    e_{(\phi)}^\phi = &1/\sqrt{g_{\phi\phi}},
\end{align}
and $\omega = -g_{t\phi}/g_{\phi\phi}$ corresponds to the frame-dragging angular velocity. This projection ensures that the initial condition rigorously satisfies the null constraint $\mathcal{H} = 0$.

\subsection{Accretion Structures}

To map the intensity of the observed light, TARTARUS models the physical matter surrounding the black hole. Currently, the code implements a geometrically thin and optically thick accretion disk based on the Novikov-Thorne model [\cite{novikov1973astrophysics}]. The disk lies in the equatorial plane, $\theta = \pi/2$, and it is bounded at its inner edge by the Innermost Stable Circular Orbit (ISCO). The time-averaged energy flux $F(r)$ emitted from the surface of the disk is computed analytically using the Page and Thorne (1974) [\cite{page1974disk}] formulation for a Kerr spacetime,

\begin{widetext}
\begin{equation}
    F(r) = \frac{3}{2 x^4 (x^3 - 3x + 2a)} \left[ x - x_0 - \frac{3a}{2} \ln \left( \frac{x}{x_0} \right) - \sum_{i=1}^{3} \frac{3(x_i - a)^2}{x_i \prod_{j \neq i} (x_i - x_j)} \ln \left( \frac{x - x_i}{x_0 - x_i} \right) \right].
\end{equation}
\end{widetext}

Here $x = \sqrt{r}$, $x_0 = \sqrt{r_{\rm ISCO}}$, $a$ is the black hole dimensionless spin parameter, and $x_1, x_2, x_3$ are the roots of the cubic equation $x^3 - 3x + 2a = 0$, defined as
\begin{equation}
    \begin{cases}
        x_1 = &2 \cos \left( \frac{\arccos a - \pi}{3} \right)\\
        x_2 = &2 \cos \left( \frac{\arccos a + \pi}{3} \right)\\
        x_3 = -&2 \cos \left( \frac{\arccos a}{3} \right)
    \end{cases}.
\end{equation}

To maintain high computational performance, this analytical expression is evaluated across a dense radial grid and interpolated  during integration using bounded lookup tables, guaranteeing efficient intensity assignments for intersecting geodesics.

\subsection{Redshift and Relativistic Beaming}

As photons travel from the accretion disk to the observer, their frequencies and intensities are shifted due to both the gravitational potential of the black hole (gravitational redshift) and the relativistic motion of the emitting plasma (Doppler). The total energy shift is quantified by the redshift factor $g$, defined as the ratio of the observed photon frequency $\nu_{\text{obs}}$ to the emitted frequency $\nu_{\text{em}}$ [\cite{cunningham1975effects}], and given by
\begin{equation}
    g = \frac{\nu_{\text{obs}}}{\nu_{\text{em}}} = \frac{(p_\mu u^\mu)_{\text{obs}}}{(p_\mu u^\mu)_{\text{em}}}.
\end{equation}

For an observer located at a very large distance from the source, its 4-velocity can be approximated as $u^\mu_{\text{obs}} = (1, 0, 0, 0)$, producing an observed energy proportional to the conserved quantity $-p_t$. The emitting plasma in the Novikov-Thorne disk is assumed to follow stable circular equatorial orbits and therefore, the 4-velocity of the emitter is $u^\mu_{\text{em}} = u^t(1, 0, 0, \Omega)$, where $\Omega = d\phi/dt$ is the Keplerian angular velocity of the disk. Using the normalization condition  $g_{\mu\nu} u^\mu_{\text{em}} u^\nu_{\text{em}} = -1$, the time component of the 4-velocity is found to be
\begin{equation}
    u^t = \frac{1}{\sqrt{-g_{tt} - 2\Omega g_{t\phi} - \Omega^2 g_{\phi\phi}}}.
\end{equation}

The redshift factor can then be explicitly written in terms of the photon's conserved canonical momenta $p_t$ and $p_\phi$ as
\begin{equation}
    g = \frac{p_t}{u^t (p_t + \Omega p_\phi)}.
\end{equation}

By Liouville's theorem, the Lorentz-invariant phase space density, $I_\nu / \nu^3$, is conserved along the photon's trajectory [\cite{rybicki1979radiative}]. Consequently, the observed bolometric intensity $I_{\text{obs}}$ of the accretion disk relates to the emitted intrinsic intensity $I_{\text{em}} \propto F(r)$ via
\begin{equation}
    I_{\text{obs}} = g^4 I_{\text{em}}.
\end{equation}

This $g^4$ amplification factor heavily modifies the visual appearance of the disk, creating the characteristic strong asymmetry due to redshift, where the side of the disk rotating toward the observer appears significantly brighter than the receding side.

\section{The TARTARUS Code} \label{sec:TheCode}

\subsection{Code Architecture}

TARTARUS is written entirely in Python 3 and it is organized around a modular scheme: the spacetime 
geometry, the emissive matter model, the observer configuration, and the 
numerical integration method are fully decoupled, allowing any component to be easily replaced or extended without modifying the others. The four primary 
modular components are:

\begin{itemize}
    \item \textbf{\lstinline|black_holes|:} Contains classes defining the 
    spacetime metric. It provides compiled functions for the metric components 
    $g_{\mu\nu}$, the geodesic right-hand side (RHS), and boundaries such as 
    the Event Horizon (EH) and ISCO. Standard implementations include 
    \lstinline|schwarzschild.py|, \lstinline|kerr.py|, and variations like 
    \lstinline|MOG_kerr.py|.
    
    \item \textbf{\lstinline|detectors|:} Defines the \lstinline|image_plane| 
    mapping, generating the initial grid of photons with impact parameters 
    $(\alpha, \beta)$ via the ZAMO tetrad formalism.
    
    \item \textbf{\lstinline|accretion_structures|:} Defines the emissive 
    matter geometry, such as the \lstinline|thin_disk|, including the 
    Novikov--Thorne flux profile $F(r)$ and the redshift factor $g$.
    
    \item \textbf{\lstinline|common.integrator|:} The numerical engine of the 
    code, which handles the progression of the affine parameter $\lambda$ via 
    one of five available integration schemes (see Section~\ref{sec:Integration Schemes and Event Detection}).
\end{itemize}

The complete module structure and the step-by-step execution flow of a 
ray-tracing run are illustrated in Figure~\ref{fig:architecture}. A typical 
run proceeds as follows: the \lstinline|detectors| module initializes the 
photon grid; \lstinline|common.integrator| evolves each null geodesic 
backward in time using the metric supplied by \lstinline|black_holes|; 
at each equatorial plane crossing, \lstinline|accretion_structures| 
evaluates the disk emissivity and redshift factor; and the resulting 
intensities are accumulated into the final image.

\subsection{Integration Schemes and Event Detection}\label{sec:Integration Schemes and Event Detection}

The numerical engine exposes all of its geodesic solvers through a single integrator interface. Five schemes are implemented natively in TARTARUS:
\begin{itemize}
    \item \textbf{Embedded Runge--Kutta pairs (RKDP45, RKCK45, RKF45):} TARTARUS implements the Dormand--Prince \cite{dormand1980rk}, Cash--Karp \cite{cashkarp1990} and Fehlberg \cite{fehlberg1969} $5(4)$ pairs. In all these methods, the fifth-order weights propagate the solution (local extrapolation) while the embedded fourth-order weights provide the error estimate that drives the adaptive step control.
    \item \textbf{Gragg--Bulirsch--Stoer \cite{gragg1965, bulirsch1966stoer}:} This method uses a modified-midpoint substepping followed by Richardson extrapolation in $h^2$, with error-controlled macro-steps.
    \item \textbf{Verlet \cite{verlet1967}:} This is a fixed-step, time-reversible symmetric-midpoint scheme that bounds the Hamiltonian drift instead of letting it grow secularly.
\end{itemize}
For the high-resolution renders of Section~\ref{sec:CodeValidation}, these schemes are compiled to thread-parallel \lstinline|python| kernels with Numba, so that the pixel generation loop bypasses the Python interpreter and runs across all cores.
Integrating backward requires strict stop conditions to optimize runtime. Each one is encoded as a smooth event function $C(\lambda, q)$ carrying a direction and a terminal/non-terminal flag, evaluated after every accepted step:
\begin{itemize}
    \item \textbf{Horizon:} This gives a terminal flag when finding the first inward crossing of $C_H = r - (r_+ + \epsilon) = 0$.
    \item \textbf{Escape:} This gives a terminal flag when finding the first outward crossing of $C_E = r - r_{\text{escape}} = 0$.
    \item \textbf{Disk:} This corresponds to a non-terminal flag. Every sign change of $C_D = \cos\theta$, i.e. each passage through the equatorial plane, is recorded to evaluate the Novikov--Thorne [\cite{novikov1973astrophysics}] disk intensity.
\end{itemize}
In all cases, a sign change of $C$ between two successive states brackets the crossing, which is then refined on the cubic-Hermite interpolant of the step, by Brent's method in the pure-Python integrators and by bisection in the Numba-compiled kernels, locating the event down to a tolerance of $\sim 10^{-12}$ in the affine parameter without shrinking the global step size.

\section{Code Validation and Performance}\label{sec:CodeValidation}
To validate the physical accuracy of the code, we perform several benchmarks evaluating both computational speed and adherence to physical conservation laws.

\subsection{Hamiltonian Constraint Conservation} \label{sec:HamiltonianConstraint}
Photon trajectories must satisfy the Hamiltonian constraint $\mathcal{H} = 0$ along the whole path. Any departure from zero is a direct, gauge-independent measure of the error accumulated by the integrator through truncation and round-off. We therefore quantify the accuracy of each integration scheme through the constraint residual,
\begin{equation}
    \epsilon_{\mathcal{H}}(\lambda) = | \mathcal{H}(\lambda) |.
\end{equation}

\begin{figure}
    \centering
    \includegraphics[width=\columnwidth]{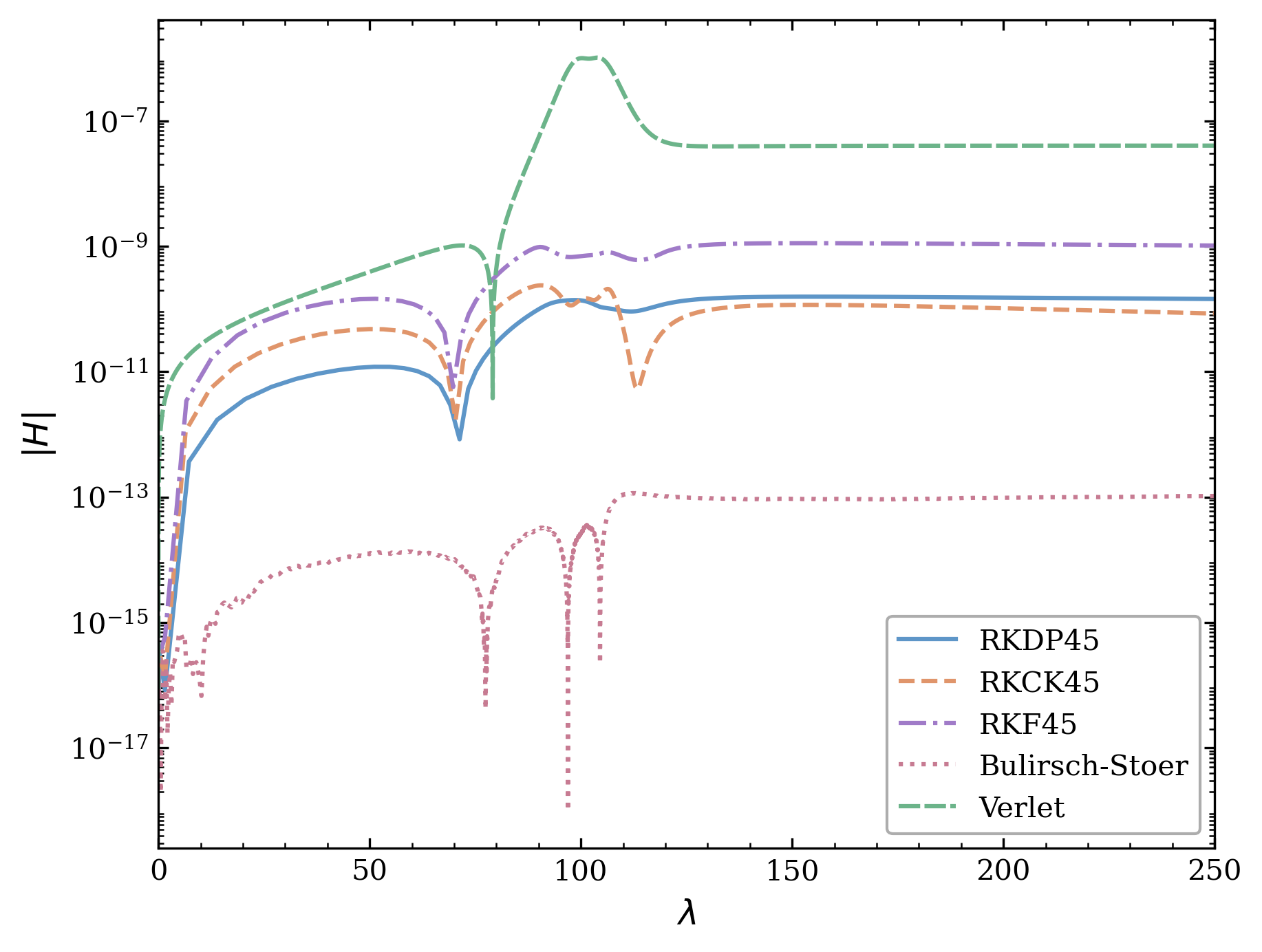}
    \caption{Hamiltonian-constraint residual $\epsilon_{\mathcal{H}} = |\mathcal{H}(\lambda)|$ along an escaping null geodesic in Schwarzschild spacetime, for the five native integrators. The photon starts at $r_0 = 100M$ with impact parameter $b = 6 > b_{\rm crit} = 3\sqrt{3}$, diving into the strong-field region and escaping to infinity. The Bulirsch--Stoer extrapolator preserves the constraint best ($\epsilon_{\mathcal{H}} \lesssim 10^{-13}$), the adaptive Runge--Kutta pairs (RKDP45, RKCK45, RKF45) hold $\lesssim 10^{-9}$, and the fixed-step Verlet method shows the largest, monotonically growing drift.}
    \label{fig:H_escape}
\end{figure}

We monitor $\epsilon_{\mathcal{H}}$ along two representative geodesics in a Schwarzschild background, both launched from $r_0 = D = 100M$ with absolute and relative tolerances of $10^{-10}$. The first geodesic corresponds to an escaping photon with an impact parameter $b = 6 > b_{\rm crit}= 3\sqrt{3}$, which dives into the strong-field region and returns to infinity. As shown in Fig.~\ref{fig:H_escape}, the five native schemes are cleanly separated by order of accuracy. The Bulirsch--Stoer extrapolator preserves the constraint best ($\epsilon_{\mathcal{H}} \lesssim 10^{-13}$), the three embedded Runge--Kutta pairs (RKDP45, RKCK45, RKF45) hold $\epsilon_{\mathcal{H}} \lesssim 10^{-9}$ along the entire trajectory, and the fixed-step Verlet method, which lacks adaptive error control, shows the largest drift. The oscillations in each curve arise from the step refinement that the adaptive solvers apply in the strong-field region near the horizon, where the gravitational field is strongest.

\begin{figure}[H]
    \centering
    \includegraphics[width=\columnwidth]{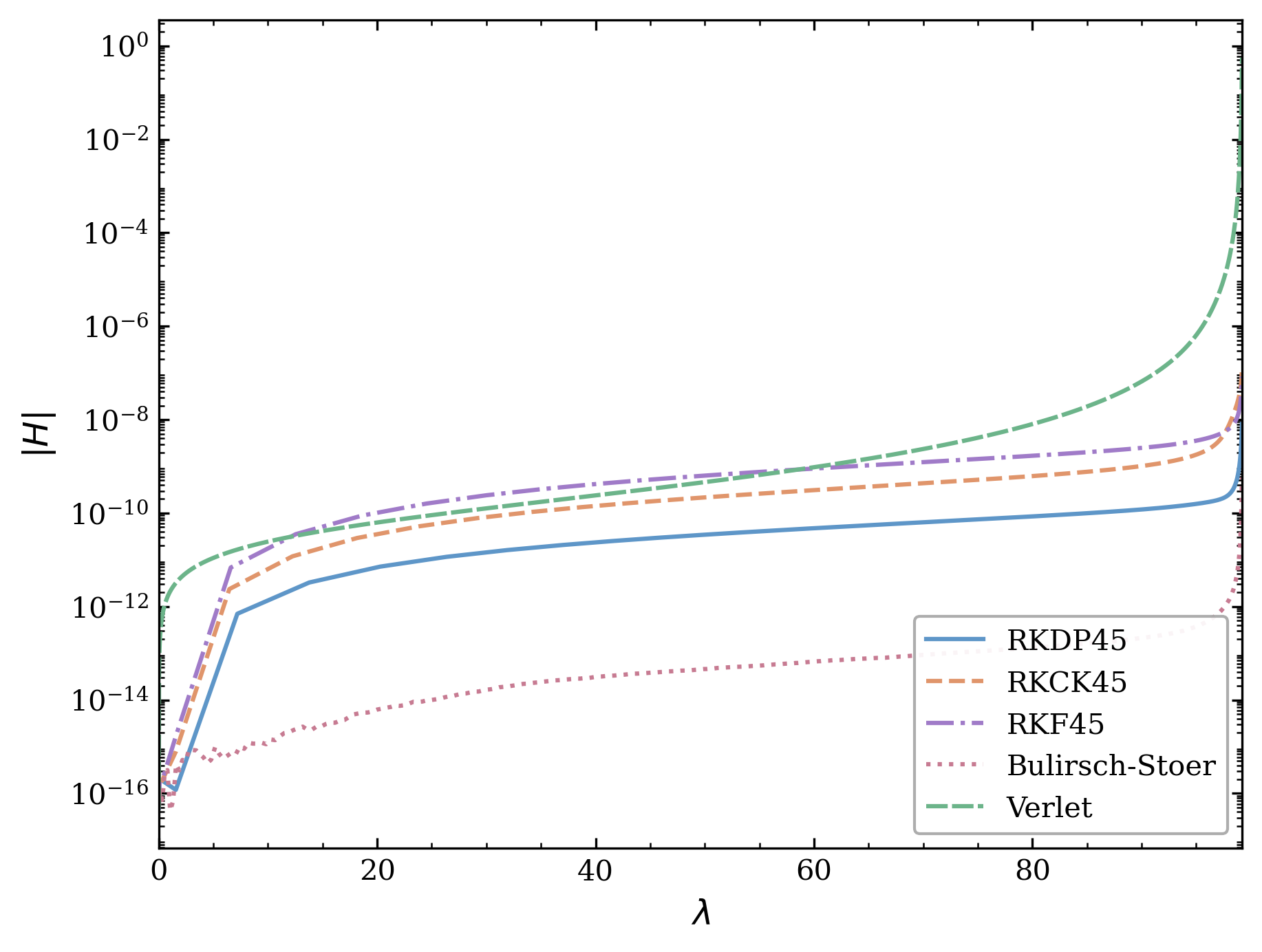}
    \caption{Same as Fig.~\ref{fig:H_escape} for a captured geodesic ($b = 3 < b_{\rm crit} = 3\sqrt{3}$), which plunges toward the event horizon. The integration is stopped at $r = 2.05M$ to avoid the coordinate singularity. Although the residual grows for every scheme, Bulirsch--Stoer remains the most accurate and Verlet the least.}
    \label{fig:H_fall}
\end{figure}

The second geodesic considered is that of a captured photon with $b = 3 < b_{\rm crit} = 3\sqrt{3}$, which plunges toward the event horizon. The corresponding constraint residual is plotted in Fig.~\ref{fig:H_fall}, showing that it grows for every scheme as the photon approaches the event horizon. We stop the integration at $r = 2.05M$ to avoid the coordinate singularity. 

In summary, the Bulirsch--Stoer extrapolator is the most accurate scheme, but the embedded Runge--Kutta pairs conserve the constraint to $\lesssim 10^{-9}$ at a small fraction of its cost. RKDP45 offers the best accuracy-to-cost balance and is therefore adopted as the production integrator for the rendering benchmarks of Section~\ref{sec:Performance}.

\subsection{Black Hole Shadows and Intensity Maps}

\begin{figure*}[p]
    \centering
    \includegraphics[width=\textwidth]{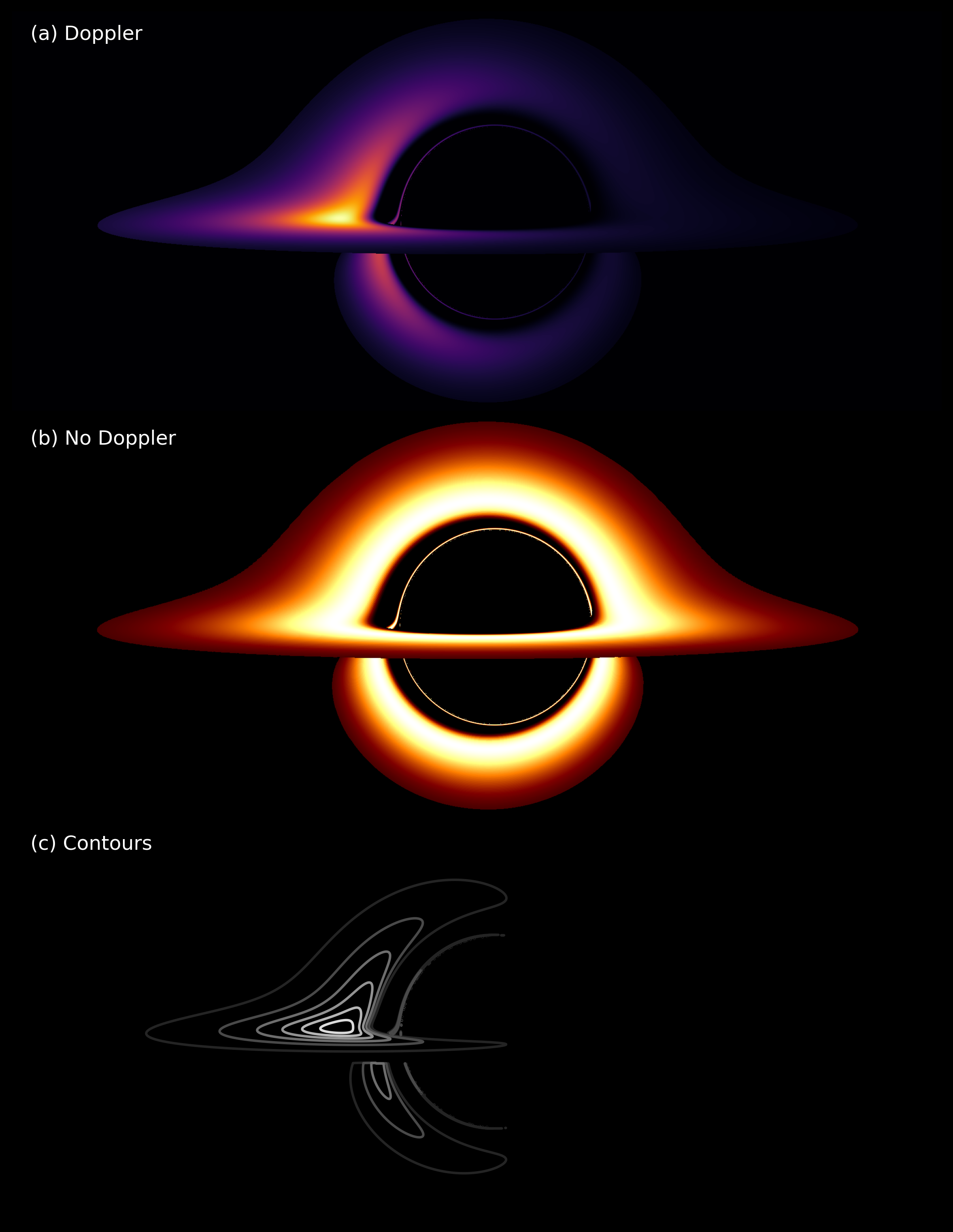}
    \caption{Image of a Kerr black hole with spin $a = 0.5$ surrounded by a Novikov--Thorne thin disk, seen by an observer at $D = 100M$ with an inclination angle $\iota = 85^\circ$. The three panels render the \emph{same} ray-traced scene in complementary ways. \textbf{(a)} Specific-intensity map including the relativistic Doppler shift: the approaching (left) side of the disk is strongly beamed and outshines the receding side, breaking the left--right symmetry. \textbf{(b)} The same map with the Doppler factor switched off, recovering the intrinsic, left--right symmetric emission of the disk together with the secondary (lensed) image seen above and below the shadow. \textbf{(c)} Iso-intensity contours of the Doppler-shifted map, tracing the brightness gradients around the photon ring and the primary/secondary disk images.}
    \label{fig:KerrUHD}
\end{figure*}

To test the rendering capabilities of the TARTARUS code, we simulated a Kerr black hole with spin parameter $a = 0.5$, surrounded by a Novikov-Thorne thin disk \cite{novikov1973astrophysics} and the result is shown in Fig. \ref{fig:KerrUHD}.

The analytical critical curve (the shadow boundary) calculated via the  \cite{bardeen1973timelike} formalism perfectly encapsulates the dark central region of the numerical output, validating the accuracy of the spatial integration.

\subsection{Performance and Benchmarking} \label{sec:Performance}
TARTARUS draws its speed from two ingredients: Numba's Just-In-Time compilation, which lifts the Python interpreter out of the inner loop, and thread parallelism, which spreads the mutually independent photon trajectories across cores. We first quantify their combined end-to-end gain over a baseline serial pure-Python implementation, then isolate the thread parallelism on its own, before benchmarking the throughput against an established external code. Unless noted otherwise, the wall-time and throughput benchmarks were run on a 16-core (32-thread) AMD Ryzen~9 8940HX laptop using all 16 cores, while the strong-scaling study used a single ppc64le (IBM POWER) cluster node.

The end-to-end gain over the original implementation is quantified in Table~\ref{tab:speedup}, which renders an image of a Schwarzschild black hole with a Novikov--Thorne thin disk ($D = 100M$) at increasing resolution. The baseline is a single-threaded, pure-Python run with the JIT disabled while the production column is the Numba-compiled kernel on 16 threads. Both execute the \emph{identical} RKDP45 algorithm, so the gap between the columns reflects the combined effect of JIT compilation and thread parallelism. Note that the production code is $\sim 500\times$ faster than the unoptimised Python baseline. The compiled kernel produces the full $1024^2$ image in $\sim 5.6$\,s and the $2048^2$ image ($\sim 4 \times 10^6$ geodesics) in just $\sim 20$\,s.

\begin{figure}[H]
    \centering
    \includegraphics[width=\columnwidth]{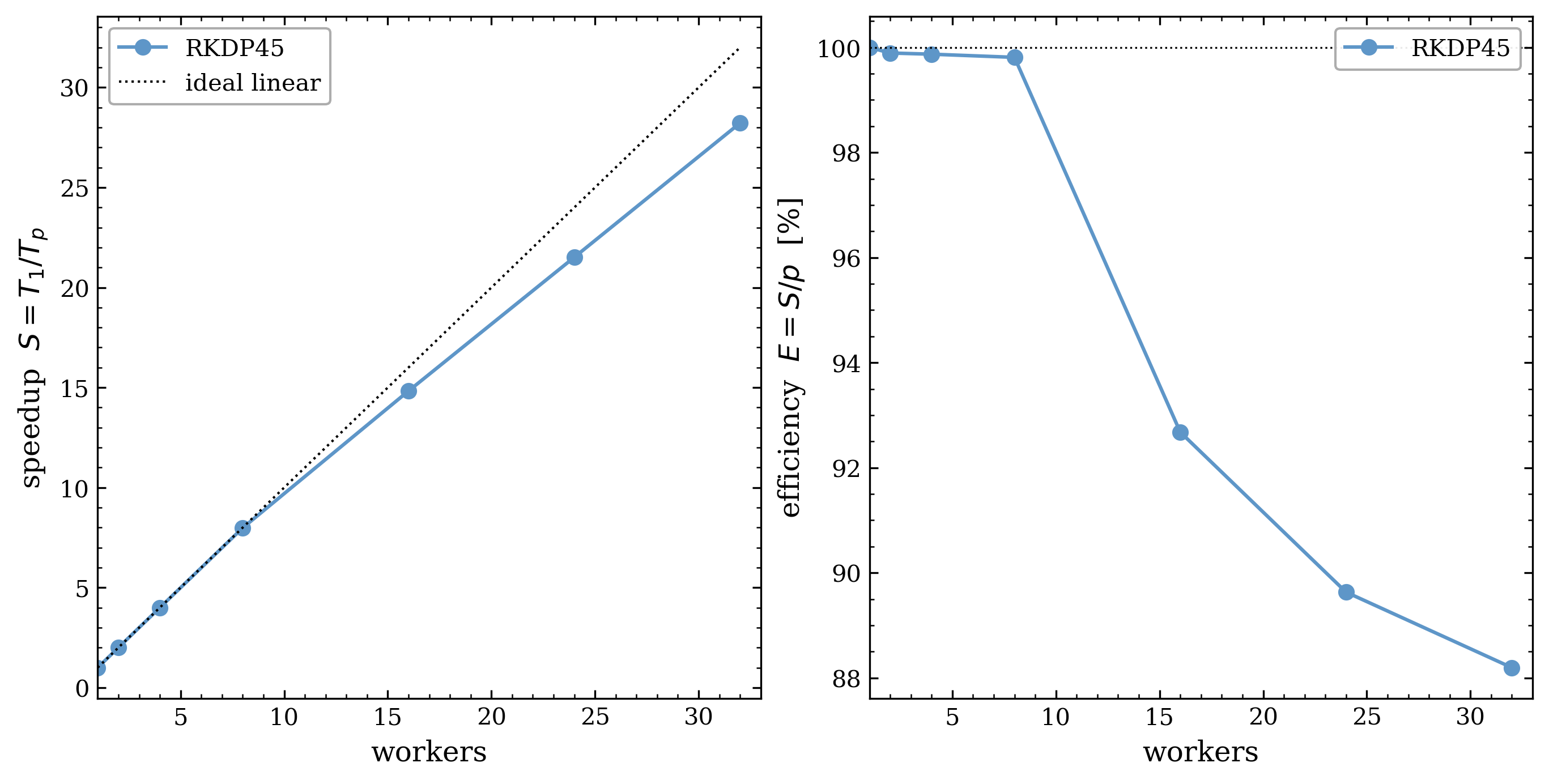}
    \caption{Strong scaling of the production RKDP45 kernel on a fixed $1024^2$ Schwarzschild shadow, on a single ppc64le (IBM POWER) cluster node (32 threads, one host). \emph{Left:} speedup $S = T_1/T_p$, the factor by which the render accelerates on $p$ threads relative to one thread (dotted line: ideal $S = p$). \emph{Right:} parallel efficiency $E = S/p$, the fraction of that ideal actually attained. The pixel loop stays close to ideal, with $E \gtrsim 90\%$ at 16 threads and $\sim 88\%$ at 32.}
    \label{fig:strong_scaling}
\end{figure}

The thread parallelism is implemented by distributing the loop over image-plane pixels with Numba's \lstinline|prange| so that each thread evolves an independent block of geodesics with no Python-level overhead. Figure~\ref{fig:strong_scaling} shows the resulting strong scaling of the RKDP45 kernel on a fixed $1024^2$ shadow: the speedup follows the ideal almost exactly, holding $\gtrsim 90\%$ parallel efficiency at 16 threads and $\sim 88\%$ across all 32, close to linear scaling.

\begin{figure}
    \centering
    \includegraphics[width=\columnwidth]{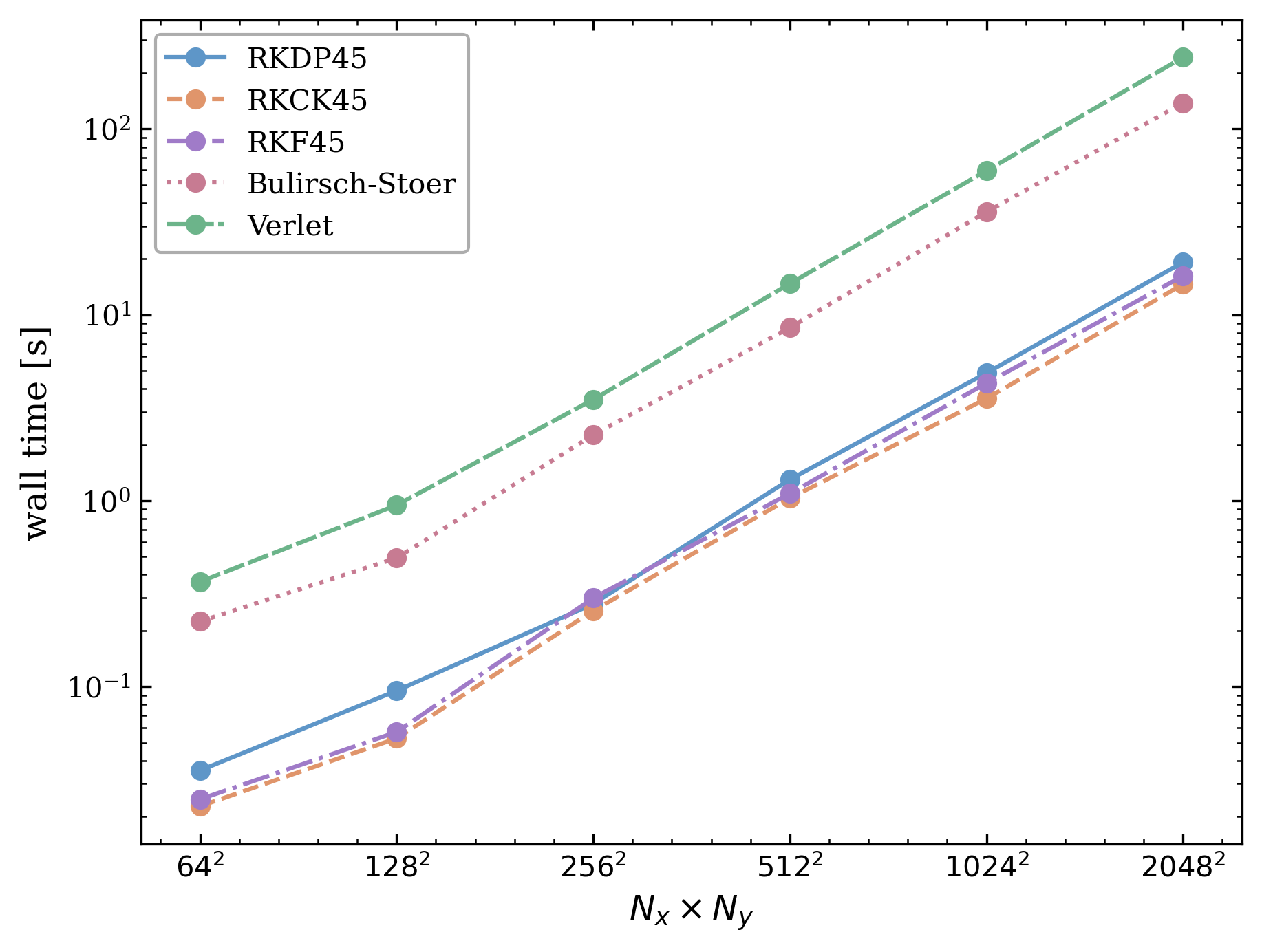}
    \caption{Wall time (logarithmic scale) versus image resolution $N_x \times N_y$ on the OSIRIS calibration scene --- a Kerr black hole with $a = 0.98$, an equatorial observer at $r_0 = 1000M$, and an image plane $[-8,8]^2$ --- for the five Numba-compiled integrators running on all 16 cores of the test machine. At $1024^2$ ($\sim 10^6$ geodesics) the adaptive Runge--Kutta kernels render the full Kerr shadow in $\sim 65$\,s, more than an order of magnitude faster than the  FORTRAN code OSIRIS on the same scene.}
    \label{fig:osiris_compare}
\end{figure}

\begin{figure}
    \centering
    \includegraphics[width=\columnwidth]{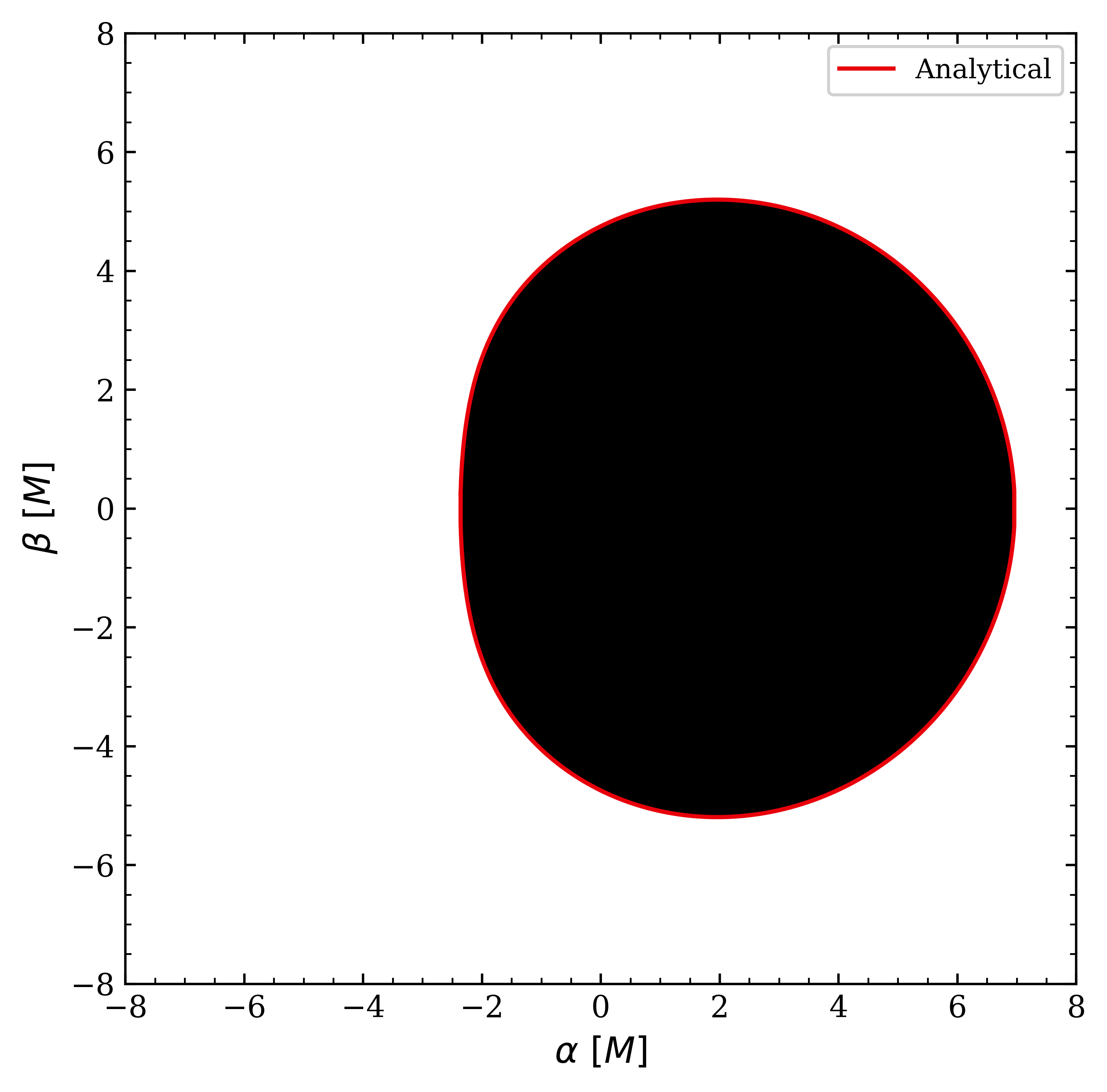}
    \caption{Kerr shadow rendered by TARTARUS for the OSIRIS calibration scene (spin $a = 0.98$, equatorial observer at $r_0 = 1000M$, image plane $[-8,8]^2$, $1024^2$ pixels, RKDP45 integrator). Backward-traced photons that reach the event horizon form the dark silhouette, while those that escape make up the light background. Frame dragging renders the silhouette asymmetric and displaced from the origin. This is the image whose wall time is reported in Figure~\ref{fig:osiris_compare}.}
    \label{fig:kerr_shadow}
\end{figure}

Combining both ingredients, we benchmark against an established code on the calibration scene of OSIRIS \cite{velasquez2022osiris}: a Kerr black hole shadow with spin $a = 0.98$, an equatorial observer at $r_0 = 1000M$, and an image plane $[-8,8]^2$ see Figure~\ref{fig:kerr_shadow}. The corresponding performance results are shown in Figure~\ref{fig:osiris_compare} plots the wall time of all five integrators against resolution. The three embedded RungeKutta pairs (RKDP45, RKCK45, RKF45) are the fastest and almost indistinguishable; the Bulirsch-Stoer extrapolator is several times slower bacause of  its many modified-midpoint sub-steps, and the fixed-step Verlet scheme is the slowest. This speed ordering mirrors the accuracy ranking of Section~\ref{sec:HamiltonianConstraint}, where Bulirsch Stoer is the most accurate and Verlet the least, and it singles out RKDP45 as the best speed/accuracy compromise. At $1024^2$ ($\sim 10^6$ geodesics) RKDP45 renders the complete Kerr shadow (Figure~\ref{fig:kerr_shadow}) in $\sim 65$\,s, against $\sim 10^3$\,s for the FORTRAN code OSIRIS on the same test. This corresponds to a performance improvement exceeding one order of magnitude when executed on standard commodity hardware and in the absence of GPU acceleration. On a hardware-independent metric, the average wall time to advance one photon by one integration step, TARTARUS reaches $\sim 35$$47$\,ns across the 16 cores, within a factor of $\sim 5$$7$ of the $7.15$\,ns of the double-precision GPU code GRay \cite{chan2013gray}. This corresponds to a performance of the same order of magnitude as a dedicated GPU.

\begin{table*}
    \centering
    \begin{tabular}{c|c|c|c|c}
         Image Resolution & Total Photons & Pure Python Time (s) & Numba JIT Time (s) & Speedup Factor \\
         \hline
         $512 \times 512$   & 262\,144    & 671.83             & 1.35  & $498\times$ \\
         $1024 \times 1024$ & 1\,048\,576 & 2\,687.3  & 5.61  & $479\times$ \\
         $2048 \times 2048$ & 4\,194\,304 & 10\,749.3 & 19.71 & $545\times$ \\
    \end{tabular}
    \caption{Serial pure-Python versus production Numba wall time for rendering an image of a Schwarzschild black hole surrounded by a Novikov--Thorne thin accretion disk (observer at $D = 100M$) at increasing resolution. The pure-Python column is a single-threaded, JIT-disabled baseline, while the Numba column is the compiled kernel on 16 threads; both execute the \emph{identical} RKDP45 algorithm, so the last column is the combined speedup from JIT compilation \emph{and} thread parallelism.}
    \label{tab:speedup}
\end{table*}

\section{Conclusions}
We have presented TARTARUS, a highly efficient, modular Python code designed for calculating null geodesics in general relativistic backgrounds. By utilizing the Numba library, TARTARUS eliminates the traditional performance bottlenecks associated with Python loops, yielding execution speeds suitable for high-resolution rendering and extensive parameter-space exploration.

The code accurately models arbitrary black hole metrics, demonstrated here using the Kerr and Schwarzschild solutions and captures physical observables such as the black hole shadow and the emission profiles of geometrically thin accretion disks. Our validation tests demonstrate excellent conservation of the Hamiltonian constraint as well as numerical renders that align perfectly with the analytical shadow boundaries.

TARTARUS provides a robust foundation for future theoretical explorations, including the modeling of optically thin relativistic plasmas, polarized radiative transfer, and ray-tracing in alternative theories of gravity. The code is structured for open-source community contributions and is highly adaptable for both educational purposes and cutting-edge astrophysical research.

Future research directions include the implementation of additional background spacetime metrics, the integration of a general relativistic magnetohydrodynamics (GRMHD) module to achieve a more accurate simulation of the accretion disk, and the development of models for selected spectral characteristics of the emitted radiation.

\section{ Code Availability}

The TARTARUS code is published as free software under the MIT license. The complete source code, alongside documentation and examples, is publicly available in the GitHub repository at \hyperlink{https://github.com/ashcat2005/TARTARUS}{TARTARUS}.

\section*{Acknowledgments}
This work was supported by the Universidad Nacional de Colombia, Hermes Grant Code 64813, and by the Research Incubator No.64 on Computational Astrophysics of the Observatorio Astronómico Nacional.




\bibliography{biblio}

\appendix
\section{Code architecture and execution pipeline}
\label{app:architecture}
\renewcommand{\thefigure}{A\arabic{figure}}
\setcounter{figure}{0}

\begin{figure*}[t]
    \centering
    \resizebox{\textwidth}{!}{%
    \begin{tikzpicture}[x=1pt, y=-1pt,
      box/.style={draw, line width=0.6pt, rounded corners=3pt, align=center, inner sep=1.5pt},
      arr/.style={-{Stealth[length=5pt,width=4pt]}, line width=1pt, draw=black!60},
      darr/.style={-{Stealth[length=5pt,width=4pt]}, line width=0.8pt, draw=black!40, dashed},
      sec/.style={font=\footnotesize, text=seccol},
      leg/.style={font=\scriptsize, anchor=west},
    ]
    \draw[seccol!60, line width=0.5pt] (20,20) -- (270,20);
    \node[sec] at (340,20) {\textsc{Code Modules}};
    \draw[seccol!60, line width=0.5pt] (410,20) -- (660,20);
    \node[box, fill=detfill, draw=detline, minimum width=148pt, minimum height=80pt, text width=144pt] at (94,72)
      {\arctitle{detectors}\\[2pt]\arcsub{image\_plane}\\\arcsub{init\_cond (ZAMO)}\\\arcsub{pixel grid $\alpha,\beta$}};
    \node[box, fill=bhfill, draw=bhline, minimum width=148pt, minimum height=80pt, text width=144pt] at (256,72)
      {\arctitle{black\_holes}\\[2pt]\arcsub{schwarzschild.py}\\\arcsub{kerr.py}\\\arcsub{MOG\_kerr.py, \dots}};
    \node[box, fill=accfill, draw=accline, minimum width=148pt, minimum height=80pt, text width=144pt] at (418,72)
      {\arctitle{accretion\_structures}\\[2pt]\arcsub{thin\_disk.py}\\\arcsub{Novikov--Thorne $F(r)$}\\\arcsub{redshift factor $g$}};
    \node[box, fill=intfill, draw=intline, minimum width=154pt, minimum height=80pt, text width=150pt] at (583,72)
      {\arctitle{common.integrator}\\[2pt]\arcsub{RKDP45, RKCK45, RKF45}\\\arcsub{Bulirsch--Stoer, Verlet}\\\arcsub{Numba prange}};
    \draw[seccol!60, line width=0.5pt] (20,130) -- (235,130);
    \node[sec] at (340,130) {\textsc{Execution Flow}};
    \draw[seccol!60, line width=0.5pt] (445,130) -- (660,130);
    \node[box, fill=detfill, draw=detline, minimum width=194pt, minimum height=86pt, text width=190pt] at (117,185)
      {\arctitle{\textcircled{\scriptsize 1} Define observer}\\[2pt]\arcsub{image\_plane($D,\iota,N_\alpha,N_\beta$)}\\\arcsub{pixel grid $(\alpha,\beta)$}\\\arcsub{$\rightarrow$ photon list}};
    \node[box, fill=bhfill, draw=bhline, minimum width=194pt, minimum height=86pt, text width=190pt] at (329,185)
      {\arctitle{\textcircled{\scriptsize 2} Initial conditions}\\[2pt]\arcsub{init\_cond($\alpha,\beta$, metric)}\\\arcsub{ZAMO tetrad $\rightarrow p^\mu$}\\\arcmath{$q_0=[x^\mu,p_\mu],\ \mathcal{H}=0$}};
    \node[box, fill=intfill, draw=intline, minimum width=194pt, minimum height=86pt, text width=190pt] at (541,185)
      {\arctitle{\textcircled{\scriptsize 3} Integrate null geodesic}\\[2pt]\arcsub{RKDP45, adaptive step}\\\arcmath{$dq/d\lambda=f(q,g^{\mu\nu})$}\\\arcsub{Numba prange (parallel)}};
    \draw[arr] (214,185) -- (232,185);
    \draw[arr] (426,185) -- (444,185);
    \draw[arr] (541,228) -- (541,240) -- (117,240) -- (117,252);
    \node[box, fill=evfill, draw=evline, minimum width=194pt, minimum height=86pt, text width=190pt] at (117,295)
      {\arctitle{\textcircled{\scriptsize 4} Event detection}\\[2pt]\arcmath{$r<r_+\ \rightarrow$ captured}\\\arcmath{$\theta=\pi/2\ \rightarrow$ disk crossing}\\\arcmath{$r>r_{\mathrm{esc}}\ \rightarrow$ escapes}};
    \node[box, fill=accfill, draw=accline, minimum width=194pt, minimum height=86pt, text width=190pt] at (329,295)
      {\arctitle{\textcircled{\scriptsize 5} Redshift and emission}\\[2pt]\arcmath{$g=p_t/u^t(p_t+\Omega p_\phi)$}\\\arcsub{$F(r)$ from interpolated table}\\\arcsub{$\rightarrow I_{\mathrm{em}}$ of the disk}};
    \node[box, fill=intfill, draw=intline, minimum width=194pt, minimum height=86pt, text width=190pt] at (541,295)
      {\arctitle{\textcircled{\scriptsize 6} Assign intensity}\\[2pt]\arcmath{$I_{\mathrm{obs}}=g^4\,F(r)$}\\\arcsub{pixel $(\alpha,\beta)\leftarrow I_{\mathrm{obs}}$}\\\arcsub{accumulate $N$ crossings}};
    \draw[arr] (214,295) -- (232,295);
    \draw[arr] (426,295) -- (444,295);
    \draw[arr] (541,338) -- (541,350) -- (340,350) -- (340,360);
    \node[box, fill=outfill, draw=outline, minimum width=400pt, minimum height=50pt, text width=396pt] at (340,385)
      {\arctitle{\textcircled{\scriptsize 7} Final rendered image}\\[2pt]\arcsub{2D intensity array $\cdot$ PNG $\cdot$ post-analysis}};
    \node[box, fill=notefill, draw=noteline, dashed, minimum width=390pt, minimum height=44pt, text width=386pt] at (215,450)
      {{\large\itshape Continuous check: $\epsilon_{\mathcal{H}}=|\mathcal{H}(\lambda)|$}\\\arcsub{evaluated at every accepted integrator step}};
    \draw[darr] (410,450) -- (430,450) -- (430,185) -- (444,185);
    \draw[seccol!50, line width=0.5pt] (20,486) -- (660,486);
    \draw[fill=detfill, draw=detline, line width=0.5pt, rounded corners=1pt] (20,494) rectangle ++(10,10);
    \node[leg] at (34,499) {detectors};
    \draw[fill=bhfill, draw=bhline, line width=0.5pt, rounded corners=1pt] (110,494) rectangle ++(10,10);
    \node[leg] at (124,499) {black\_holes};
    \draw[fill=accfill, draw=accline, line width=0.5pt, rounded corners=1pt] (212,494) rectangle ++(10,10);
    \node[leg] at (226,499) {accretion};
    \draw[fill=intfill, draw=intline, line width=0.5pt, rounded corners=1pt] (295,494) rectangle ++(10,10);
    \node[leg] at (309,499) {common.integrator};
    \draw[fill=outfill, draw=outline, line width=0.5pt, rounded corners=1pt] (430,494) rectangle ++(10,10);
    \node[leg] at (444,499) {output};
    \draw[fill=notefill, draw=noteline, dashed, line width=0.5pt, rounded corners=1pt] (492,494) rectangle ++(10,10);
    \node[leg] at (506,499) {internal check};
    \end{tikzpicture}}
    \caption{Modular architecture and end-to-end backward ray-tracing pipeline of TARTARUS. \emph{Top:} the four code modules. \emph{Bottom:} the seven-step execution flow, from defining the observer's image plane, through the backward integration of null geodesics, event detection, redshift and emission, and intensity assignment, to the final rendered image. The dashed box marks the continuous Hamiltonian-constraint check $\epsilon_{\mathcal{H}} = |\mathcal{H}(\lambda)|$ evaluated at every accepted integrator step.}
    \label{fig:architecture}
\end{figure*}
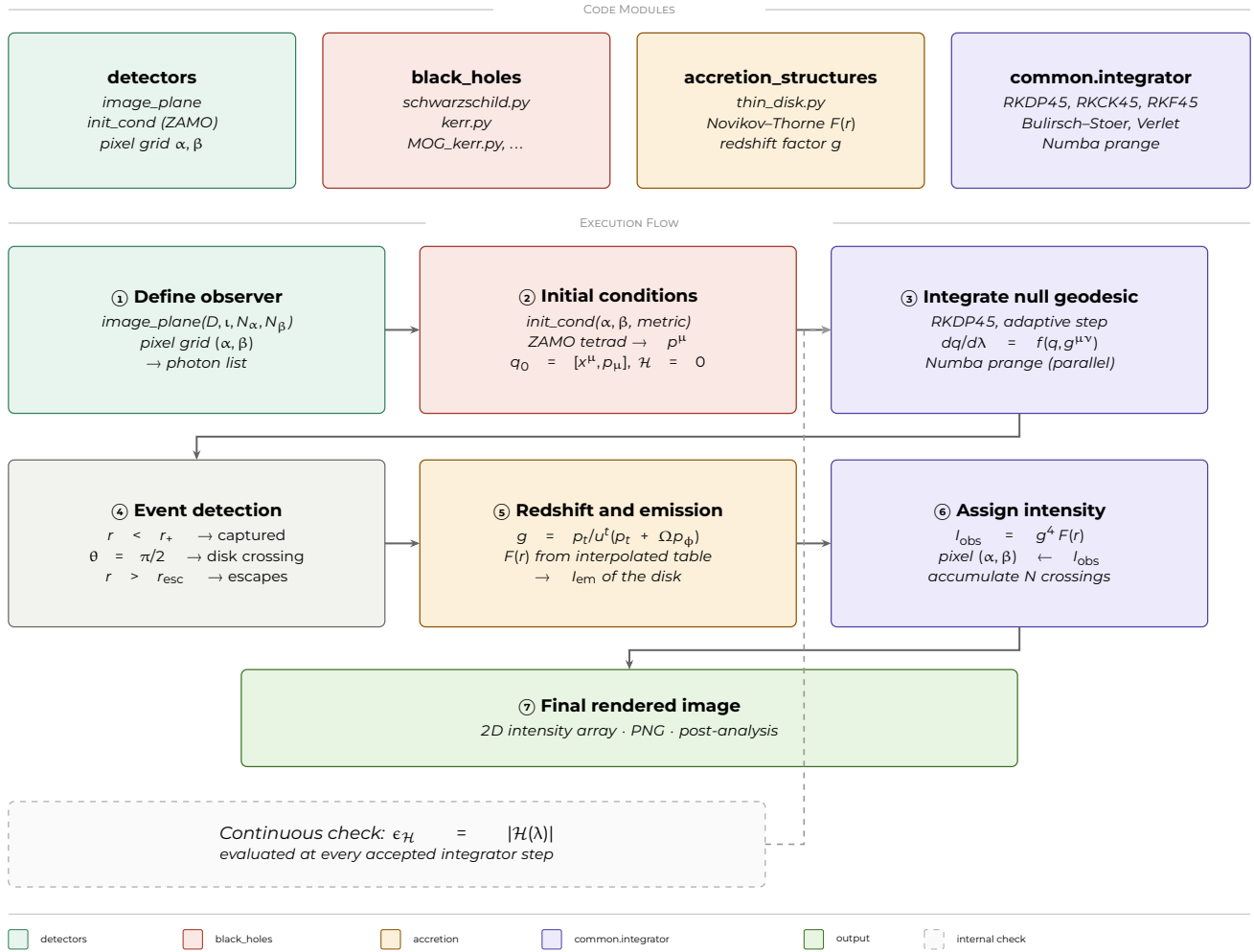

Figure~\ref{fig:architecture} schematically summarizes the modular architecture of TARTARUS and the end-to-end backward ray-tracing pipeline described in Section~\ref{sec:TheCode}, from the observer's image plane to the final rendered image.

\end{document}